\def\ts     {\thinspace}
\def\kms    {\ifmmode{{\rm \ts km\ts s}^{-1}}\else{\ts km\ts s$^{-1}$}\fi}
\def\msol     {\ifmmode{{\rm M}_{\odot}}\else{M$_{\odot}$}\fi}
\def\etal   {{\rm et\ts al.}}
\def\aco  {{\ifmmode{{^{12}{\rm CO}(J\!=\!1\! \to \!0)}}\else{{$^{12}{\rm CO}(J\!=\!1\! \to \!0)$}}\fi}}
\def\bco  {\ifmmode{^{12}{\rm CO}(J\!=\!2\! \to \!1)}\else{$^{12}{\rm CO}(J\!=\!2\! \to \!1)$}\fi}
\def\m    {\ifmmode{\mu {\rm m}}\else{$\mu$m}\fi}
\def\cco  {\ifmmode{^{13}{\rm CO}(J\!=\!1\! \to \!0)}\else{$^{13}{\rm CO}(J\!=\!1\! \to \!0)$}\fi}
\def\dco  {\ifmmode{^{13}{\rm CO}(J\!=\!2\! \to \!1)}\else{$^{13}{\rm CO}(J\!=\!2\! \to \!1)$}\fi}
\def\eco  {\ifmmode{{\rm C}^{18}{\rm O}\-(J\!=\!1\! \to \!0)}\else{${\rm C}^{18}{\rm O}\-(J\!=\!1\! \to \!0)$}\fi}
\def\nh   {\ifmmode{N(\hi)}\else{$N$(\hi)}\fi}
\def\hun    {\ifmmode{I_{100}}\else{$I_{100}$}\fi}
\def\sex    {\ifmmode{I_{60}}\else{$I_{60}$}\fi}
\def\hh     {\ifmmode{{\rm H}_2}\else{H$_2$}\fi}
\def\nhh     {\ifmmode{N({\rm H}_2)}\else{$N$(H$_2$)}\fi}
\def\zwco   {\ifmmode{^{12}{\rm CO}}\else{$^{12}{\rm CO}$}\fi}
\def\nzwco   {\ifmmode{N(^{12}{\rm CO})}\else{$N(^{12}{\rm CO})$}\fi}
\def\wzwco   {\ifmmode{W(^{12}{\rm CO})}\else{$W(^{12}{\rm CO})$}\fi}
\def\drco   {\ifmmode{^{13}{\rm CO}}\else{$^{13}{\rm CO}$}\fi}
\def\ndrco   {\ifmmode{N(^{13}{\rm CO})}\else{$N(^{13}{\rm CO})$}\fi}
\def\wdrco   {\ifmmode{W(^{13}{\rm CO})}\else{$W(^{13}{\rm CO})$}\fi}
\def\tex    {\ifmmode{T_{ex}({\rm CO})}\else{$T_{ex}({\rm CO})$}\fi}
\def\ha     {\ifmmode{{\rm H}\alpha}\else{${\rm H}\alpha$}\fi}

\newcommand{\hi}{H\,{\small{\sc I}}}

\documentclass[here]{aa}

\usepackage{graphics}

\begin{document}

\hyphenation{con-sti-tu-ents there-by mar-gin-al-ly stat-is-tics}

\thesaurus{01(09.02.1, 11.09.1 M\,82, 11.09.4, 11.11.1, 11.19.3, 13.25.4)}

\title{Evidence for an Expanding Molecular Superbubble in M\,82\\}

\author{Axel Wei\ss
\and 
Fabian Walter
\and
Nikolaus Neininger
\and 
Uli Klein
}

\institute{Radioastronomisches Institut der Universit\"at Bonn, Auf dem H\"ugel 71, 53121 Bonn, Germany
}

\offprints{A.Wei\ss, aweiss@astro.uni--bonn.de}
\date{Received 23 February 1999 / Accepted 06 April 1999}
\authorrunning{A. Wei\ss\ et al.}
\maketitle


\begin{abstract}

We present evidence for an expanding superbubble in M\,82 (diameter:
$\approx$\,130\,pc, expansion velocity: $\approx$\,45\,\kms, mass: 
$\approx\,8\cdot10^6\,$\msol). It is seen in the
{$^{12}{\rm CO}\\(J\!=\!1\! \to \!0)$}, {\bco}, {\cco}\ and {\eco}\ lines. 
The superbubble is centred around the most powerful supernova 
remnant, 41.9+58. The CO
observations show that the molecular superbubble already broke out
of the disk. This scenario is supported by ROSAT HRI observations
which suggest that hot coronal gas originating from inside the
shell is the main contributor to the diffuse X--ray outflow in
M\,82.  We briefly discuss observations of the same region at other
wavelengths (radio continuum, optical, \hi, X--rays, ionized gas). 
From our spectral line observations, we derive a kinematic
age of about $10^6$ years for the superbubble.  Using simple
theoretical models, the total energy needed for the creation of this
superbubble is of order $2\times10^{54}\,$ergs. The required energy 
input rate (0.001 SN\,yr$^{-1}$) is reasonable given the 
high supernova (SN) rate of $\approx 0.1$ 
SN\,yr$^{-1}$ in the central part of M\,82. As much as 10\% of 
the energy needed to create the superbubble is still present 
in form of the kinetic energy of the expanding molecular shell.
Of order 10\% is conserved in the hot X--ray emitting gas 
emerging from the superbubble into the halo of M\,82.
This newly detected expanding molecular superbubble is believed to be
powered by the same objects that also lie at the origin of the prominent
X--ray outflow in M\,82. It can therefore be used as an alternative
tool to investigate the physical properties of these sources.


\end{abstract}

\section{Introduction}

M\,82 is the best studied nearby starburst galaxy 
(${\rm D}\,=\,3.25\, {\rm Mpc}$). The central few hundred
parsecs of this galaxy are heavily obscured by dust and gas which 
hides the central starburst region against direct observations at optical
wavelengths. Evidence for strong star--forming activity in the
central region comes from radio (e.g. Kronberg \etal\
\cite{kronberg81}) and infrared observations (e.g. Telesco \etal\ 
\cite{telesco91}) and also from the prominent bipolar outflow visible
in \ha\ (e.g. Bland \& Tully \cite{bland88}, McKeith \etal\ \cite{mckeith95}, 
Shopbell \& Bland--Hawthorn \cite{shopbell98}) and in X-rays 
(e.g. Bregman \etal\ \cite{bregman}). 
The massive star formation (SF) is believed to be fuelled by the large 
amount of molecular gas which is present in the centre of M\,82. 

On the other hand, SF effects the distribution and kinematics of the
surrounding interstellar medium (ISM). Recent millimetre 
continuum observations (Carlstrom \& Kronberg 
\cite{carlstrom90}) suggested that the  H{\small II} regions in M\,82
have swept up most of the surrounding neutral gas and dust into dense
shells. This is in agreement with the standard picture: shells are 
created by young star--forming regions through strong stellar 
winds of the most massive stars in a cluster and through subsequent 
type--II supernovae (e.g.\ Tenorio--Tagle \& Bodenheimer \cite{tenorio88}).  
These processes are thought to blow huge cavities filled with coronal 
gas into their ambient ISM (e.g.\ Cox \& Smith \cite{cox74}, Weaver \etal\
\cite{weaver77}).  This hot interior is then believed to drive the
expansion of the outer shell of swept--up material.

Once superbubbles reach sizes that are comparable to the thickness of
a galaxy's disk, the bubble will eventually break out into the
halo. This then leads to an outflow of the hot gas with 
velocities much higher than the expansion of the shell within the 
disk of the galaxy. In the following we present 
evidence for a molecular superbubble in M\,82 which already broke out 
of the disk and seems to contribute significantly to the 
well--known prominent outflow of M\,82.

\section{Observations} 

\subsection{Molecular Lines}

For our analysis we used the \aco\ data cube obtained by Shen \& Lo
(\cite{shen}) with the BIMA array (spatial resolution: $2.5''$) and
the \cco\ data cube from Neininger \etal\ (\cite{nico13}) observed
with the Platau de Bure interferometer (PdBI) (spatial resolution:
$4.2''$). In addition, we used unpublished PdBI data of the
\bco\ and \eco\ transitions. These observations where carried out in 
April 1997 in the CD configuration, resulting in a spatial resolution 
of $1.4''\times 1.2''$ (\bco) and $3.7'' \times 3.5''$ (\eco), and a velocity
resolution of 3.3 \kms\ and 6.8 \kms, respectively.  In order to
increase the sensitivity to extended CO emission we combined the
\bco\ data cube with single dish measurements obtained with the IRAM 30m
telescope. These observations were carried out in May
1998. The combination is essential for this particular study
because the receding part of the superbubble is only marginally visible
in the mere interferometer maps. A full account of the data
reduction will be given elsewhere.

\subsection{Evidence for an Expanding Superbubble}

\begin{figure}[h]
\hspace*{0.5cm}
\resizebox{7cm}{!}{\includegraphics{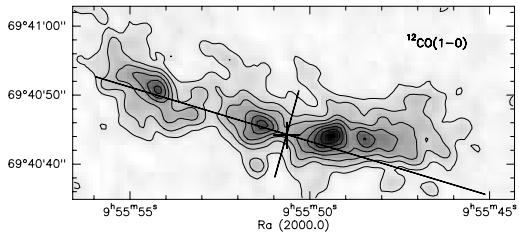}}
\caption{Integrated $^{12}{\rm CO}$ line emission from Shen \& Lo (1995). 
The lines along the major and minor axis of M\,82 indicate the
orientations of the pv--diagrams shown in Figs.~\ref{pv} and
\ref{pv-minor}. The cross marks the position of SNR 41.9+58. }
\label{12co10}
\end{figure}

\begin{figure}[h]
\vspace*{-0.4cm}
\hspace*{0.5cm}
\resizebox{6.5cm}{!}{\includegraphics{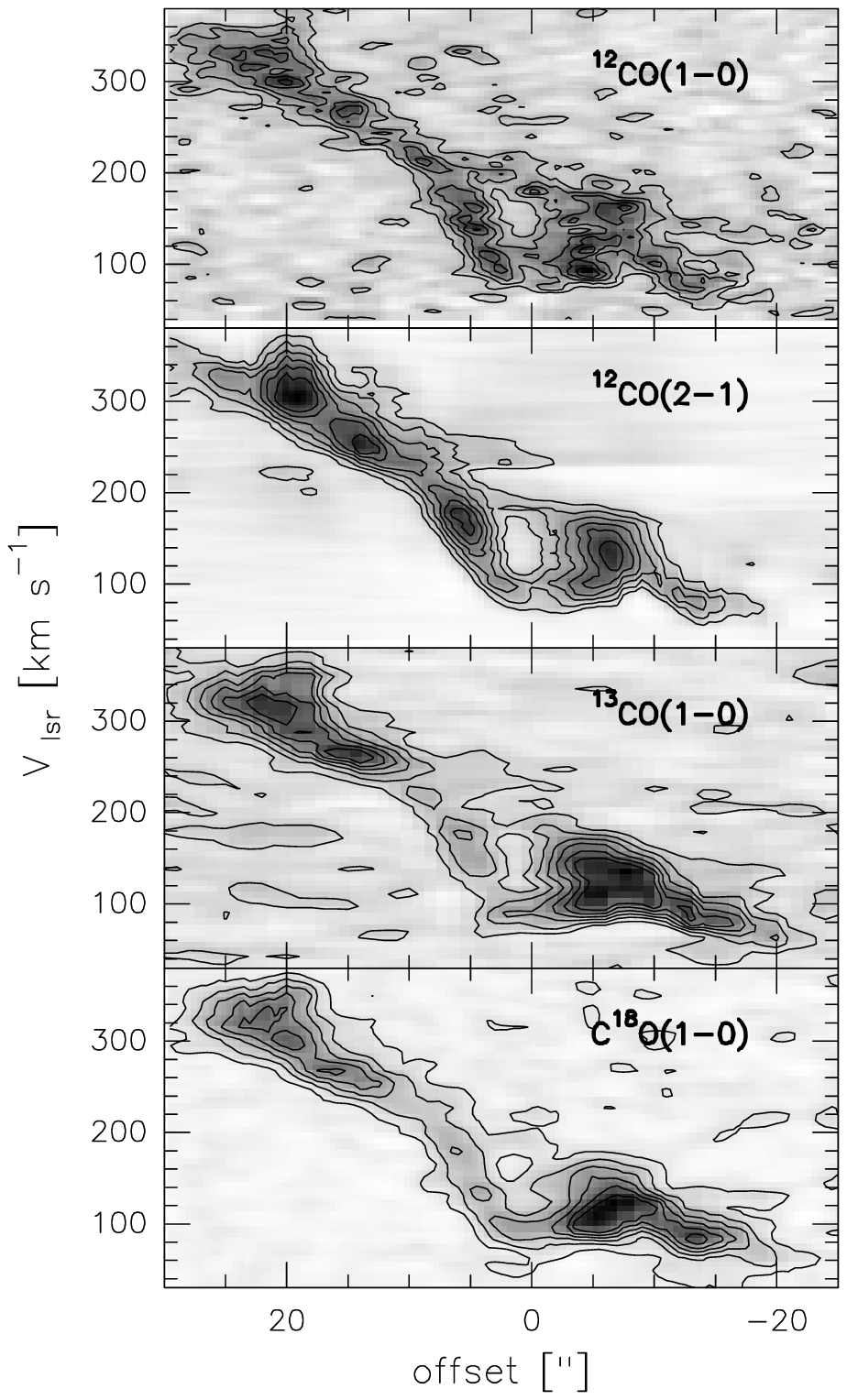}}
\caption{The pv--diagrams along the major axis of M\,82 (orientation 
shown in Fig.~\ref{12co10}) for the \aco, \bco, \cco\ and \eco\ cubes. 
The slice is centred on SNR 41.9+58.}
\label{pv}
\end{figure}

Fig.~\ref{12co10} shows the integrated $^{12}{\rm CO}$ line emission
published by Shen \& Lo (\cite{shen}). The cross corresponds to the
position of the supernova remnant 41.9+58 (SNR 41.9+58) which
is the strongest cm continuum point--source in M\,82 (Kronberg \etal\
\cite{kronberg81}). It is considered to be the aftermath of a
'hypernova', which exhibits a radio luminosity 50--100 times greater
than typical for type--II SNe (Wilkinson \& de Bruyn
\cite{wilkinson}). The line along the major axis indicates the
orientation of the position--velocity (pv) diagrams shown in
Fig.~\ref{pv}.  

\begin{figure}[h]
\hspace*{-0.5cm}
\resizebox{8cm}{!}{\includegraphics{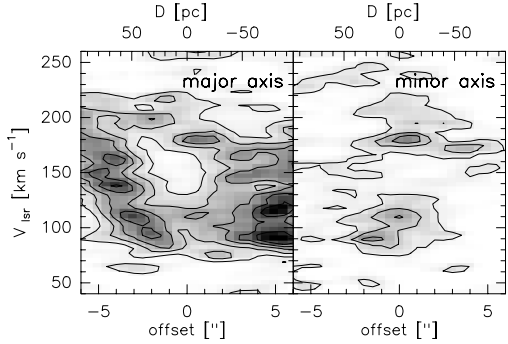}}
\caption{The pv--diagrams for the \aco\ transition  along the 
major and minor axis of M\,82. The spatial axis for each 
diagram is centred on SNR 41.9+58. The pv--diagram along the major
axis is an enlargement of Fig.~\ref{pv} (top).  
The orientation of the cut along the minor axis is indicated in 
Fig.~\ref{12co10}.}

\label{pv-minor}
\end{figure}

\noindent The pv--cut is orientated in such a way that the signature of
the expanding superbubble is visible most distinctly. The angular axis 
of the pv diagrams correspond to the offset in arcseconds from SNR
41.9+58. Besides a constant velocity gradient (cf. Shen \& Lo
\cite{shen}) the pv--diagrams show an expanding ring--like feature 
centred on SNR 41.9+58, with a central velocity 
of ${\rm V}_{{\rm lsr}}\! \approx \! 150\,\kms$.  The approaching 
velocity component of the ring is
clearly seen in all cubes and is centred at $100\, \kms$.  The receding
component is only marginally visible in the $^{13}{\rm CO}\\(J\!=\!1\! \to \!0)$ 
and \eco\ lines; its
central velocity is about $190\, \kms$. An enlargement of the pv--diagram
along the major axis is shown in Fig.~\ref{pv-minor} (left). From this
dia\-gram we estimate the radius of the ring to be ($65\,\pm\, 5$)~pc and
the expansion velocity ($45\, \pm \, 5$)~\kms.  A pv--diagram along the minor
axis centred on SNR 41.9+58 is presented in Fig.~\ref{pv-minor}
(right). The orientation of the cut is shown in Fig.~\ref{12co10}. 

\noindent Fig.~\ref{pv-minor}\,(right) reveals two emission features 
centred at the position of SNR 41.9+58, which correspond to the approaching 
($v \! \approx \! 100\,\kms$) and
the receding component ($v\! \approx \! 190\,\kms$) of the expanding
superbubble. Hardly any CO emission with a velocity between
$v \! = \! 100\, \kms$ and $v \! = \! 190\, \kms$ is found south 
and north of SNR 41.9+58. The pv--diagrams therefore show 
that the expanding molecular shell has already broken out of the disk 
and now only shows the signature of an expanding molecular ring.
It should be noted that the remaining molecular gas in M\,82
shows clear solid--body rotation.

\subsection{Other wavelengths}

\paragraph{Radio continuum observations:\\}
Radio continuum observations at 408 MHz by Wills \etal\
(\cite{wills97}) unveiled a prominent `hole' of approximately 100 pc
diameter around SNR 41.9+58 which they attribute to free--free
absorption. They propose that this feature is due to absorption by a
large H\,{\small{\sc II}} region that has been photoionized by a
cluster of early--type stars of which the progenitor of SNR 41.9+58
was originally a member. It should be noted, however, that SNR 41.9+58 
is only about 50 years old (Wilkinson \& de Bruyn \cite{wilkinson}) 
and therefore cannot be the source that drives the expansion 
of the superbubble. But its presence supports the scenario that the 
expanding molecular superbubble is powered by an interior stellar cluster.

\vspace*{-0.3cm}

\vspace*{-1cm}
\paragraph{Neutral hydrogen (H{\small I}):\\}
\hi\ emission line studies show that the central kpc of M\,82
is dominated by absorption (Yun \etal\ \cite{yun93}), which makes
it difficult to study the
\hi\ kinematics in the central part of the galaxy.  Recent \hi\
absorption studies against supernova remnants in M\,82 by Wills \etal\
(\cite{wills98}) disclosed two absorption features against SNR
41.9+58. The velocities of the components are $87 \pm 7\, \kms\ $ and
$200 \pm 7\, \kms$. The \hi\ component at 87 \kms\ can be
attributed to the approaching component of the expanding superbubble. 
The absorption feature at 200 \kms, if associated with the receding part,
seems to contradict the hypothesis that SNR 41.9+58 is located within
the expanding superbubble. However, the velocity resolution of the
absorption study was rather poor (26.4 \kms) and other explanations 
cannot be ruled out at this point.

\paragraph{Optical observations:\\}

Optical observations of the centre of M\,82 suffer heavily 
from light absorption by the prominent dust lanes, rendering an 
optical analysis of this particular region impossible. 
HST V-- and I--band images of the centre of M\,82 (O'Connell \etal\ 
\cite{connell95}) do not reveal any optical sources in the 
vicinity of SNR 41.9+58.  The prominent \ha\ outflow is visible north
and south of the absorbing dust lane (Bland \& Tully \cite{bland88},
McKeith \etal\ \cite{mckeith95}, Shopbell \& Bland--Hawthorn 
\cite{shopbell98}). The orientation 
of its filaments suggests that the \ha\ outflow emerges at least partly
from the location of the molecular superbubble.

\paragraph{X--ray observations:\\}

\begin{figure}[h]
\hspace*{0.3cm}
\resizebox{8cm}{!}{\includegraphics{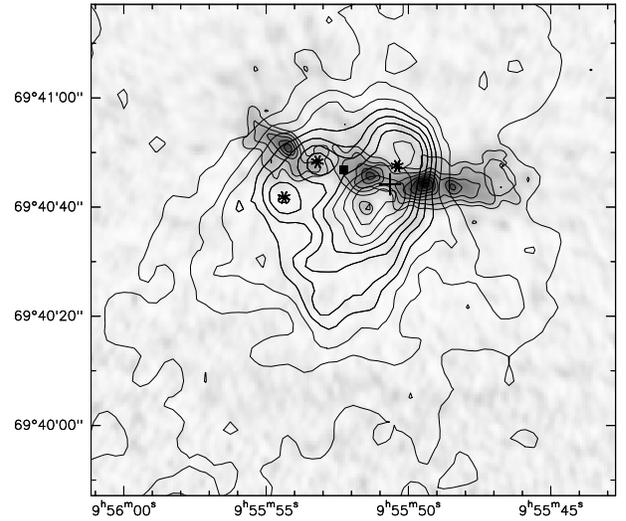}}
\caption{Integrated \aco\ line emission of M\,82 (greyscale). 
The contour map represents the diffuse X--ray emission without the
strongest point--source in the core of M\,82 as observed by the ROSAT
HRI.  The three X-ray point--sources are marked with a star, the
position of SNR 41.9+58 by a cross; the centre of M\,82 ($2.2\,\m\ $
peak) is indicated by a filled box.}
\label{xray}
\end{figure}

Fig.~\ref{xray} shows an overlay of archival ROSAT HRI X--ray data 
(Bregman \etal\ \cite{bregman}) onto the CO emission. To emphasize
the diffuse, extended X--ray outflow we subtracted a model of the
brightest X--ray point source at $\alpha=9^h55^m50^s.4$,
$\delta=69^{\circ}40'47.5''$ (J2000.0) and smoothed the residual image
to a $6''$ resolution.  The positions of the three point sources 
in the field are marked with stars, SNR 41.9+58 is marked by a 
cross (see also Bregman \etal).  The overlay shows that most of the diffuse
X--ray emission arises from the vicinity of SNR 41.9+58. The shape of
the diffuse emission is elongated, extending several arcminutes along
the minor axis of the galaxy. Model calculations by Bregman \etal\ show 
that the emission is consistent with an outflow of heated material from the
central region coincident with the position of the expanding molecular
superbubble. 

\noindent We estimate an upper limit for the energy 
still present in the hot gas emerging from the superbubble. For the
calculation we adopt a density ${\rm n}=0.2\,{\rm cm}^{-3}$ and 
a temperature ${\rm kT}=1\,{\rm keV}$ for the central region 
(Bregman \etal) as an upper limit for the mean values in the outflow. 
We find an upper limit for the energy of 
${\rm E}_{\rm gas} \le 2.7 \times 10^{53}\,$ergs present in a 
cylindrical volume with $r=130\,{\rm pc}$ and $z=550\,{\rm pc}$ centred 
on the superbubble. From the thickness of the molecular disk of M\,82 
($\approx 90\,{\rm pc}$) and the expansion velocity of the superbubble 
we estimate a time elapsed since outbreak of 
$\tau\,\approx\,3 \times 10^5\,{\rm yr}$.
The required outflow velocity to reach
the observed height above the disk therefore is 
${\rm v}\,\approx\,1000\,\kms$. This value is comparable with 
estimates based on particle aging as
derived from radio continuum observations 
(Seaquist \etal\ \cite{seaquist85}) and measured outflow
velocities in \ha\ (McKeith \etal\ \cite{mckeith95}).

\paragraph{Ionized gas:\\}
 
Most studies of the ionized gas in M\,82 unfortunately do not provide 
sufficient spatial and spectral resolution to reveal details of the region 
under study. A high--resolution study of the  Ne\,{\small{\sc II}} 
emission line has been published by Achtermann \& Lacy (\cite{achtermann}). 
The integrated line emission in the south--western
region of M\,82 shows a similar double--peaked feature centred on SNR
41.9+58 as traced by the CO. The pv--cut along the major axis clearly shows
a disturbed velocity field in the vicinity of SNR 41.9+58.
The approaching side of the expanding superbubble at 95 \kms\ is clearly
visible while the receding side at 190 \kms\ is only marginally seen. 
The diameter of the Ne\,{\small{\sc II}} shell ($\approx$\,100\,pc) 
seems to be somewhat smaller than the corresponding CO 
feature, suggesting that the inner 
part of the expanding superbubble is ionized. Similar features are 
visible in the distribution and the kinematics of 
the H41$\alpha$ emission line (Seaquist \etal\ \cite{seaquist96}).  
The Ne\,{\small{\sc II}} distribution might represent the transition 
from the hot coronal gas with a temperature of several $10^{6}\,$K 
towards the cold molecular rim of the superbubble.

\section{Discussion}

The picture that emerges from the observations presented above is the
following: a major SF event at the centre of what today shows up as the
prominent expanding molecular superbubble created a cavity 
filled with coronal gas by the combined effects of
strong stellar winds and SN explosions. The pressure of the hot--gas
interior drove the expansion of the shell of swept--up material
until it broke out of the disk.  The subsequent outflow of hot
material today shows up as diffuse X--ray emission and contributes to 
the prominent \ha\ filaments.

From the observational parameters of the expanding superbubble
(expansion velocity and diameter) we estimate a kinematic age of the
superbubble of $1\times10^{6}$ years. The kinematic age is an
upper limit for the actual age since the superbubble is most probably
decelerating. We use Chevalier's equation (Chevalier \cite{chev74})
to derive the amount of energy needed to create the expanding superbubble.
Since Chevalier's equation applies to \hi\ shells only we corrected
the energy input by a factor of 2 to correct for the difference in
mass between \hh\ and H.  We estimate an ambient \hh\ density prior to
the creation of the shell of order $120\,{\rm cm}^{-3}$. This is done
by converting the \aco\ line integral to \hh\ column density using a
conversion ratio of N(\hh)/W(CO) = $1.2 \times 10^{20}$ cm$^{-2}$
K$^{-1}$ km$^{-1}$s (Smith \etal\ \cite{smith}) and estimating the
volume of the region from which material was swept up to be
cylindrical with $z=100\,{\rm pc}$ and $r=65\,{\rm pc}$.  We derive a
total energy of order $2\times 10^{54}\,$ergs, which corresponds to an
energy equivalent of approximately 1000 type--II SNe and the strong
stellar winds of their progenitors. With the estimate for the age of
the superbubble this leads to a SN rate of 0.001 SN yr$^{-1}$ for the
central stellar cluster. This is a reasonable number given the fact
that the SN rate in the central part of M\,82 was estimated to be
about 0.1 SN yr$^{-1}$ (Kronberg \etal\ \cite{kronberg81}). The 
total \hh\ mass of the superbubble is about $8\times 10^6\,\msol$. 
Less than 15\% of the total energy is still present in the hot gas 
which emerges from the superbubble; the fraction of kinetic 
energy of the molecular superbubble is about 10\% 
(${\rm E}_{\rm kin}\,\approx\,1.6 \times 10^{53}\,$ergs). 
It should be noted, however, that the numbers given above are only order of
magnitude estimates.

The wealth of observations presented above indicates that the region
under study represents an unusually active segment of the starburst in
M\,82.  The finding of an expanding molecular superbubble in this
particular region supports this view and provides clear evidence for a
footprint of violent SF in the ISM of M\,82. Furthermore the analysis of the
molecular superbubble presents an alternative tool to investigate the
energy release of the central source which drives the X--ray and
contributes to the \ha\ outflow.

\acknowledgements{A.W. and F.W. acknowledge DFG grant III GK--GRK 118/2. 
We thank J. Shen and K.Y. Lo for making available their CO data,
J. Kerp for his help on the ROSAT data and the referee,
P. Kronberg, for helpful comments. We acknowledge the IRAM 
staff for carrying out the observations and the help provided 
during the data reduction.}

\end{document}